\documentclass[12pt]{iopart}
\usepackage{iopams,setstack}
\usepackage{amstext,graphicx,color}
\usepackage[compress]{cite}

\newcommand{\TR}{\text{Tr}}
\newcommand{\mydagger}{{\dagger}}
\newcommand{\phdagger}{\phantom{\mydagger}}
\newcommand{\tb}{{\bar t}}
\newcommand{\ket}[1]{|#1\rangle}
\newcommand{\bra}[1]{\langle#1|}
\newcommand{\expval}[1]{\langle#1\rangle}
\newcommand{\CC}{\mathcal{C}}
\newcommand{\TC}{\text{T}_{\CC}}

\newcommand{\tmax}{t_{\text{max}}}
\newcommand{\convz}{\ast}
\newcommand{\ret}{{\text{r}}}
\newcommand{\adv}{{\text{a}}}
\newcommand{\mats}{{\text{\tiny M}}}
\newcommand{\tv}{{\makebox{$\neg$}}}
\newcommand{\vt}{{\reflectbox{$\neg$}}}
\newcommand{\les}{<}
\newcommand{\lar}{>}
\newcommand{\vect}[1]{{\bi #1}}
\newcommand{\Vk}{{\vect{k}}}
\newcommand{\Vq}{{\vect{q}}}
\newcommand{\rtime}{\tau}
\newcommand{\ramp}{\Delta\kappa}
\newcommand{\exE}{\Delta E}
\newcommand{\quench}{\text{quench}}
\begin{document}

  \title[Near-adiabatic parameter changes
  in correlated systems]{Near-adiabatic parameter changes
    in correlated systems: Influence of the ramp protocol on the excitation energy}

  \author{Martin Eckstein$^{1,2}$, Marcus Kollar$^1$}
  
  \address{$^1$
    Theoretical Physics III,
    Center for Electronic Correlations and Magnetism,
    Institute of Physics,
    University of Augsburg,
    86135 Augsburg, 
    Germany}

  \address{$^2$
    Institut of Theoretical Physics,
    ETH Zurich, 
    Wolfgang-Pauli-Str. 27,
    8093 Zurich, 
    Switzerland}

  \date{November 6, 2009}

  \begin{abstract}
    We study the excitation energy for slow changes of the hopping
    parameter in the Falicov-Kimball model with nonequilibrium
    dynamical mean-field theory.  The excitation energy vanishes
    algebraically for long ramp times with an exponent that depends on
    whether the ramp takes place within the metallic phase, within the
    insulating phase, or across the Mott transition line.  For ramps
    within metallic or insulating phase the exponents are in agreement
    with a perturbative analysis for small ramps. The perturbative
    expression quite generally shows that the exponent depends
    explicitly on the spectrum of the system in the initial state and
    on the smoothness of the ramp protocol.  This explains the
    qualitatively different behavior of gapless (e.g., metallic) and
    gapped (e.g., Mott insulating) systems.  For gapped systems the
    asymptotic behavior of the excitation energy depends only on the
    ramp protocol and its decay becomes faster for smoother ramps. For
    gapless systems and sufficiently smooth ramps the asymptotics are
    ramp-independent and depend only on the intrinsic spectrum of the
    system. However, the intrinsic behavior is unobservable if the
    ramp is not smooth enough.  This is relevant for ramps to small
    interaction in the fermionic Hubbard model, where the intrinsic
    cubic fall-off of the excitation energy cannot be observed for a
    linear ramp due to its kinks at the beginning and the end.
  \end{abstract}

  \pacs{71.27.+a,  67.85.-d}

  % \maketitle

  \section{Introduction}
  
  In equilibrium thermodynamics, adiabatic processes are defined as
  quasistatic processes without heat exchange with the environment.
  The entropy remains constant during an adiabatic process, while it
  always increases if the process takes place in a finite time  and is 
  therefore no longer quasistatic and reversible. These fundamental 
  concepts are closely related to the adiabatic theorem of quantum 
  mechanics~\cite{Born1928,Kato1950,Avron1999} for an isolated
  system which evolves according to the Schr\"odinger
  equation
  \begin{equation}
    i\hbar\ket{\psi(t)}
    =
    H(t)\ket{\psi(t)},
  \end{equation}
  with a time-dependent Hamiltonian $H(t)$, i.e., a system that is
  subject to external fields or to changes of its parameters, but not coupled to heat or
  particle reservoirs. The adiabatic theorem states that a system that
  is initially in the ground state evolves to the new ground state
  during an infinitesimally slow change of the Hamiltonian, whereas it
  cannot follow a parameter change that takes place in a finite time,
  resulting in a non-zero excitation energy.  The paradigm for this
  crossover from adiabatic to nonadiabatic behavior in a quantum
  system is the exactly solvable Landau-Zener  model 
  \cite{Landau1932a,Zener1932a}, i.e., a two-level system
  $H_{LZ}(t)$ $=$ $vt \sigma_z + \gamma \sigma_x$ that is driven
  through an avoided level crossing with finite speed $v>0$
  ($\sigma_z$ and $\sigma_x$ are Pauli matrices). When the system is
  in the ground state $\ket{\phi_0(-\infty)}$ $=$ $(1,0)^\text{+}$ at
  time $t$ $=$ $-\infty$, the probability to find the system in the
  excited state $\ket{\phi_1(\infty)}$ $=$ $(1,0)^\text{+}$ at time
  $t$ $\to$ $\infty$ vanishes exponentially when the speed $v$ is
  small compared to the scale $\gamma^2/\hbar$ set by the gap 
  $\gamma$ at the avoided crossing,
  $|\expval{\psi(t\to\infty)|\ket{\phi_1(\infty)}}|^2$ $\sim$
  $\exp(-\pi\gamma^2/v\hbar)$.

  The above Landau-Zener formula can be generalized to various multilevel cases
  \cite{Brundobler1993a,Shytov2004a,Volkov2004a,Dobrescu2006a}, from which,
  e.g., the demagnetization probability for the transverse-field Ising model was
  obtained~\cite{Ostrovsky2006a}. However, 
  for correlated systems in general the
  Landau-Zener results cannot be directly applied, because 
  essentially all matrix elements of an interacting many-particle Hamiltonian change 
  in a complicated way
  upon variation of  one of its parameters.
  The investigation of slow changes of external parameters in correlated systems
  has recently received considerable attention due to its 
  relevance for experiments with ultracold atomic gases in optical lattices~\cite{Bloch2008a}, 
  in which quantum-many body systems can be kept under well-controlled conditions.
  In those systems time-dependent control of the parameters is not only
  of practical importance (as discussed below), but it also allows to test 
  fundamental theoretical predictions. For example, the Landau-Zener result was
  indeed experimentally confirmed in a Bose-Einstein 
  condensate loaded into an accelerated optical lattice~\cite{Zenesini2009a}.

  Various slow parameter changes in many-body systems have recently
  been studied \cite{Barouch1970,Polkovnikov2005a,Zurek2005a,Dziarmaga2005a,Barankov2006a,Damski2006a,Wubs2006a,Dziarmaga2006a,Cucchietti2007a,Klich2007a,Altland2008a,Polkovnikov2008b,Barankov2008a,Sen2008a,Tomadin2008a,Itin2009a,Grandi2009a,Divakaran2009a,Moeckel2009b}.
  For a general {\em ramp} the system 
  is initially in the ground state and some parameter of the Hamiltonian is 
  then changed to a new value within a time interval $\rtime$, either linearly 
  or nonlinearly with time. To investigate the crossover from the extreme 
  nonadiabatic limit $\rtime=0$ (i.e., a  sudden {\em quench} of the Hamiltonian) 
  to possibly adiabatic behavior in the limit $\rtime$ $\to$ $\infty$ a measure 
  for the degree of nonadiabaticity is needed. A popular quantity for this
  purpose is the excitation energy $\Delta E(\rtime)$ after the ramp,
  i.e.,
  \begin{equation}
    \label{exe}
    \exE(\rtime) = E(\rtime) - E_0(\rtime),
  \end{equation}
  where $E_0(\rtime)$ is the ground-state energy of the Hamiltonian after
  the ramp. For initial states at non-zero temperature, the entropy 
  increase provides a more natural measure of nonadiabaticity in 
  general.  However, entropy is uniquely defined only for thermal equilibrium, 
  and thus it can only be computed after the ramp is complete and 
  the system has thermalized. On the other hand, isolated many-body
  systems do not necessarily thermalize quickly after changes in the
  Hamiltonian~\cite{Heims65,Girardeau1969,Sengupta2004a,Cazalilla2006a,Rigol2007a,Kollath2007,Manmana07,Eckstein2008a,Moeckel2008a,Kollar2008,Rigol2008a,Rossini2009b,Barmettler2009a,Eckstein2009a},
  in particular for integrable systems,
  as demonstrated experimentally with ultracold 
  gases~\cite{Kinoshita2006a}. 
  In contrast to the entropy the internal energy
  is always well-defined, regardless of whether the system passes through a series
  of thermal or nonthermal states in the limit of a quasistationary
  process.

  In the present work we only consider systems that are initially in 
  the ground state. If the excitation energy $\Delta E(\rtime)$ vanishes 
  in the limit of  long ramp times, $\rtime$ $\to$ $\infty$, the system is considered
  to behave adiabatic.  It is expected that the excitation energy is
  still small for finite ramp times $\rtime$, just as in the Landau-Zener 
  formula, when the ground state is protected by a gap
  for all parameters throughout the ramp~\cite{Polkovnikov2008b}.  
  However, the excitation energy is not exponentially small 
  ($\Delta E(\rtime)$ $\propto$ $\exp(\text{const}/\rtime)$) in general.
  As we will show below, the asymptotic decrease of $\Delta E(\rtime)$ 
  for large $\tau$ can depend both on the intrinsic properties of the 
  many-body system and on the ramp protocol. In particular for gapped
  systems the ramp protocol can be used to make 
  $\Delta E(\rtime)$ arbitrarily small, but in general it
  often vanishes only algebraically. This behavior is known from the 
  Landau-Zener model, where the excitation is exponentially small 
  only when the avoided level crossing is traversed from $t$ $=$ $-\infty$ 
  to $t$ $=$ $+\infty$, whereas the excitation probability is proportional 
  to $1/\rtime^2$  and hence much larger if the evolution takes place 
  from $t=0$ to $t=\infty$, i.e., starting exactly at the center of the level
  crossing~\cite{Vitanov1999a,Damski2006a}.

  The situation is completely different for gapless systems, such as
  the exactly solvable one-dimensional transverse-field Ising model in
  which the gap vanishes at exactly one value of the transverse field.
  When the magnetic field is ramped across this critical point the
  excitation energy
  is~\cite{Polkovnikov2005a,Zurek2005a,Dziarmaga2005a}
  \begin{equation} 
    \label{deltae-tau}
    \Delta E(\rtime) \sim \rtime^{-\eta},~~(\rtime\to\infty)
  \end{equation} 
  with a rational exponent $\eta$ $=$ ${\textstyle\frac12}$. Similar
  results were obtained for a number of other quantum critical
  systems, such as the Bose-Hubbard model~\cite{Cucchietti2007a} or
  the random field Ising model~\cite{Dziarmaga2006a}.  However, the
  existence of a quantum critical point is not a necessary condition 
  to obtain a nonanalytic relation $\Delta  E(\rtime)$~\cite{Polkovnikov2008b,Altland2008a}.
  Equation~(\ref{deltae-tau}), with various values of the exponent $\eta$,
  holds for ramps within gapless phases of several gapless
  systems~\cite{Polkovnikov2008b}. For a continuous bath of harmonic
  oscillators, which model the low-energy excitations of a large class of 
 systems,  the exponent $\eta$ for a slow squeeze of the oscillator mass
  depends on the spatial dimension~\cite{Polkovnikov2008b}: An
  analytic relation $\Delta E(\rtime)$ $\sim$ $\rtime^{-2}$ is found
  for all dimensions $d$ $\ge$ $3$, while $\eta$ is noninteger for $d$
  $=$ $2$. For $d=1$, the thermodynamic limit does not commute with
  the limit of large $\rtime$, i.e., the prefactor in
  Eq.~(\ref{deltae-tau}) increases with system
  size~\cite{Polkovnikov2008b}, suggesting that adiabatic behavior is
  impossible for that class of one-dimensional systems.

  The excitation energy during a nonadiabatic ramp, and its dependence
  on the ramp duration $\rtime$ is not only a fundamental property of 
  a quantum many-body system, but it is also of practical interest 
  for experiments with cold atomic gases. Various ramping procedures are
  used in experiment to transform one phase into another, and the
  available time for the process cannot be too long in order to avoid
  extrinsic losses. On the other hand, whether theoretical predictions 
  are actually observable in experiment can depend in a subtle way on 
  the unavoidable excitation during the preparation of the
  state~\cite{Barankov2006a,Reischl2005a,Rey2006a,Ho2007a,Yoshimura2008a,Cramer2008a,Pollet2008a}. 
  When the ramp duration is  fixed to a given maximum value, it thus
  becomes important to find the {\em optimal ramp} through which a 
  given point in parameter space  can be reached through minimal excitation 
  of the system~\cite{Barankov2008a}.  In general, it
  is plausible that any additional term in the Hamiltonian should be
  switched on slowly, so as to build up the correlations that it
  favors without incurring to high energy cost, and increasing the
  speed at later times. 

  In view of these issues the question arises to what extent the 
  dependence of the excitation energy on the ramp duration $\rtime$ is 
  determined by intrinsic properties of the system, and to what extent it
  is influenced by the details of the ramp. In this paper we give a perturbative argument 
  that holds in the limit of small ramp amplitudes and allows to separate
  an intrinsic contribution to the excitation energy and a ramp shape 
  dependent contribution. In some cases the latter can mask the intrinsic 
  contribution such that the behavior of the excitation energy in the limit 
  of long ramp times $\rtime$ is completely determined by the ramp shape.
  Furthermore, we present results for the excitation of the Falicov-Kimball
  model after various ramps. In this model, which can be solved exactly using 
  nonequilibrium dynamical mean-field theory (DMFT),  Eq.~(\ref{deltae-tau}) 
  is found to hold  with an exponent $\eta$ that is different for ramps across 
  the metal-insulator transition, within the metallic phase, and within the 
  insulating phase. Our numerical results for $\eta$ in this model support the 
  scenario obtained from the perturbative argument.
  
  The paper is organized as follows. In Sec.~\ref{secfiniteramps} we
  show results for the excitation energy in the Falicov-Kimball model
  in nonequilibrium dynamical mean-field theory.  In
  Sec.~\ref{smallramp-sec-derivation} we develop a perturbative
  argument for small ramp amplitudes 
  and discuss the implications for gapped and
  gapless systems, such as the metallic and Mott insulating phases of
  the Falicov-Kimball and the fermionic Hubbard model. A conclusion in
  Sec.~\ref{concl} closes the presentation.

  \section{Ramps in the Falicov-Kimball model}
  \label{secfiniteramps}

  \subsection{Model}

  Below we present results for the excitation energy $\Delta
  E(\rtime)$ for ramps of different types in the Falicov-Kimball model
  \cite{Falicov1969a}, with Hamiltonian
  \begin{eqnarray}%
    \label{FKM}
    H_{\text{Falicov-Kimball}}(t)
    =&
    \sum_{ij} V_{ij}(t)\,c_{i}^{\mydagger} c_{j}^{\phdagger}
    + U(t) \sum_{i} n_{i}^f n_{i}^c
    \nonumber\\
    &-\mu\sum_{i} n_{i}^c-(\mu-E_f)\sum_in_{i}^f.
  \end{eqnarray}%
  Here $c_{i}^{(\mydagger)}$ and $f_{i}^{(\mydagger)}$ are
  annihilation (creation) operators for the itinerant and immobile
  electrons, respectively, and $n_{i}^c$ $=$ $c_{i}^{\mydagger} c_{i}$
  ($n_{i}^f$ $=$ $f_{i}^{\mydagger} f_{i}^{\phdagger}$) are their
  local densities.  Hopping between sites $i$ and $j$, with amplitude
  $V_{ij}(t)=V(t)t_{ij}$, is possible only for the mobile $c$
  particles.  Note that although the $f$ electrons are immobile,
  the equilibrium state of $H_{\text{Falicov-Kimball}}$ does not
  correspond to one quenched $f$ configuration but rather to a state
  with annealed disorder, where each $f$ state contributes according
  to the free energy of the $c$ particles.

  In the context of dynamical mean-field theory (DMFT) \cite{Georges1996a}, which becomes
  exact in infinite dimensions \cite{Metzner1989a}, the
  Falicov-Kimball model has a long history because it can be mapped
  onto a solvable single-site  problem \cite{Brandt1989a,vanDongen1990a,vanDongen1992a,Freericks2003a}.
  A Mott metal-insulator transition occurs at a critical interaction
  $U_c$ for half-filling (at density $n_c$ $=$ $n_f$ $=$ $\frac12$),
  as well as a transition to a charge-ordered state at sufficiently
  low temperatures. The physics of the Falicov-Kimball model thus partly
  resembles that of its parent, the fermionic Hubbard model,
  \begin{equation}
    \label{HM}
    H_{\text{Hubbard}}(t) =\sum_{ij,\sigma=\uparrow,\downarrow}
    V_{ij}(t) c_{i\sigma}^{\mydagger} c_{j\sigma}^{\phdagger}
    + U(t) \sum_{i} n_{i\uparrow} n_{i\downarrow}
    - \mu \sum_{i,\sigma=\uparrow,\downarrow}n_{i\sigma},
  \end{equation}
  with two mobile spin species.

  DMFT can be applied to nonequilibrium situations
  \cite{Schmidt2002a,Freericks2006a,Freericks2008b,Eckstein2008a,Tran2008a,
  Joura2008a,Tsuji2008a,Tsuji2009a,Eckstein2009a,Eckstein2009b},
  in which case the effective single-site problem for the Falicov-Kimball
  model is still quadratic and can be solved using equations of motion
  \cite{Freericks2006a}. Here we extend the exact solution of the 
  Falicov-Kimball model for an interaction quench \cite{Eckstein2008a} 
  to a numerical solution that can be applied to arbitrary time dependencies 
  in $V(t)$ and $U(t)$ (\ref{app1}). This allows us to study the excitation 
  after ramps of the hopping or the interaction strength.  We employ a 
  set of hopping amplitudes for which the density of states has a 
  semielliptic shape,
  \begin{eqnarray}
    \label{rho-semi}
    \rho(\epsilon)
    =
    \frac{1}{L}\sum_{\bi{k}}\delta(\epsilon-\epsilon_{\bi{k}})
    =
    \frac{1}{2\pi}\,\sqrt{4-\epsilon^2},
  \end{eqnarray}
  where $\epsilon_{\bi{k}}$ are the eigenvalues of the hopping matrix
  $t_{ij}$ and $L$ is the number of lattice sites. Furthermore, we consider
  only the homogeneous phase at half-filling, for which the chemical potential 
  is fixed at $\mu$ $=$ $U/2$ and the $f$-orbital energy at $E_f$ $=$ $0$.  
  In this case the critical interaction for the equilibrium Mott transition is
  $U_c=2V$~\cite{vanDongen1990a}.

  \subsection{Linear ramp protocol}

  We consider linear ramps in the Falicov-Kimball model~(\ref{FKM}) in
  DMFT for the homogeneous paramagnetic phase at half-filling.  We
  assume that the system is in the ground state for times $t$ $<$ $0$.
  For $0\le t \le \rtime$ the hopping parameter $V$ is changed
  according to the ramp protocol
  \begin{equation}
    \label{vramp}
    V(t) =
    \left\{
      \begin{array}{ll}
        V_i & t \le 0 \\
        V_i + \Delta V \, r(t/\rtime) & 0<t<\rtime\\
        V_f=V_i+\Delta V & t \ge \rtime\,,
      \end{array}
    \right.
  \end{equation}
  where $V_i$ is the initial hopping amplitude, $\rtime$ is the total
  ramp time, $\Delta V$ is the ramp amplitude, and $r(x)$ is the ramp
  shape. The latter is a monotonously increasing function with $r(0)$
  $=$ $0$ and $r(1)$ $=$ $1$.  We set the energy scale by $V_i$
  $\equiv V\equiv 1$, so that time is measured in units of $1/V$.
  (From now on we set $\hbar=1$.)
  The energy of the system per lattice site is given by
  \begin{eqnarray}
    \label{etot-def}
    E(t)
    &\equiv
    \frac{1}{L}\left[
      V(t) \sum_{ij} t_{ij}(t) \expval{c^{\mydagger}_i(t) c^{\phdagger}_j(t) }
      +
      U(t)
      \sum_{i} \expval{c^{\mydagger}_i(t) c^{\phdagger}_j(t) f_i^\mydagger(t) f_i^{\phdagger}(t)}
    \right].
  \end{eqnarray}
  The excitation after a ramp is then obtained from the
  difference~(\ref{exe}), where $E_0(\rtime)$ is the
  energy~(\ref{etot-def}) of the ground state of the Hamiltonian after
  the ramp. The DMFT solution for ramps in $V(t)$ (and also $U(t)$) is
  described in \ref{app1}.

\begin{figure}[t]
    \centerline{\includegraphics[width=0.85\columnwidth]{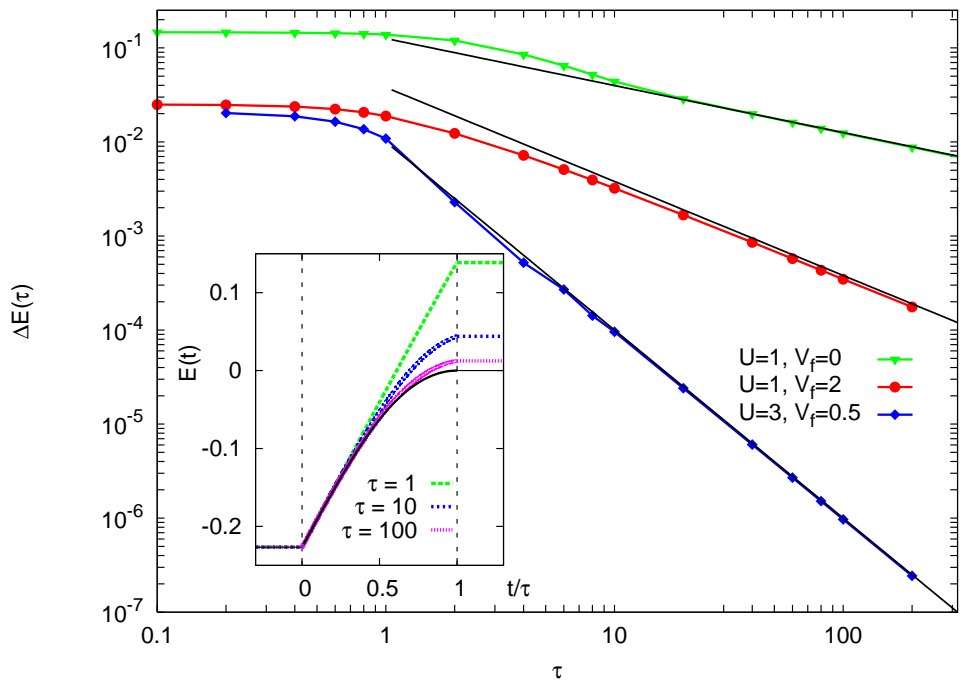}}
    \caption{Excitation energy~(\ref{exe}) after linear ramps of the
      hopping parameter [Eq.~(\ref{vramp}), $r(x)=x$] within the
      metallic phase ($U=1$, $V_f=2$), within the insulating phase
      ($U=3$, $V_f=0.5$), and across the metal-insulator transition
      ($U=1$, $V_f=0$).
      The energy scale is set by  $V_i$ $\equiv$ $V=1$. The
      curves become independent of $\rtime$ in the quench regime
      $\rtime$ $\lesssim$ $1/V$.  The solid black lines, with a 
      slope $1/2$, $1$, and $2$ (from top to bottom), correspond 
      to the asymptotic behavior~(\ref{exe-fkm}).
      Inset: Internal energy $E(t)$ [Eq.~(\ref{etot-def})] during ramps
      (\ref{vramp}) of the hopping amplitude in the Falicov-Kimball model
      ($r(x)=x$, $U=1$, and $V_f=0$), using various ramp durations $\rtime$. 
      For $t<0$ and $t>\rtime$, the energy is constant.
      The solid black line is the internal energy $E(t)$ in the
      ground state at $U$ $=$ $1$ and hopping $V(t)$.}
    \label{fig-exe1}
  \end{figure}

  The time evolution of the energy~(\ref{etot-def}) is plotted
  in the inset of Fig.~\ref{fig-exe1}
  during a ramp~(\ref{vramp}) with linear profile $r(x)=x$.
  For small ramp-durations ($\rtime$ $=$ $1$), the energy rises
  linear with time. In this case, the system is essentially quenched,
  i.e., its state $\ket{\psi(t)}$ remains unchanged during the ramp,
  and the energy is thus only determined by the ramp protocol, $E(t)$
  $\approx$ $\expval{\psi(0)|H(t)|\psi(0)}$.  In the opposite limit
  $\rtime\to\infty$, the energy adiabatically follows the ground-state
  energy $E_0(t)$ for hopping parameter $V(t)$ (solid line in inset of
  Fig.~\ref{fig-exe1}), in accordance with the adiabatic theorem.

  We now focus on the excitation $\Delta E(\rtime)$ after the ramp,
  which is plotted in Fig.~\ref{fig-exe1} for linear
  ramps~(\ref{vramp}) within the gapless metallic phase ($U=1$,
  $V_f=2$), within the gapped insulating phase ($U=3$, $V_f=0.5$), and
  across the metal-insulator transition ($U=1$, $V_f=0$), which occurs
  in the equilibrium system at $U$ $=$ $1$ and $V=0.5$. From Fig.~\ref{fig-exe1}
  one can estimate the crossover timescale $\tau_\quench$, which
  separates the regime in which the state of the system cannot follow
  the parameter change ($\rtime<\tau_\quench$) from the
  \emph{adiabatic regime} in which $\Delta E(\rtime)$ decreases with
  increasing ramp-duration $\rtime$ ($\rtime>\tau_\quench$).
  Independent of the ramp parameters, $\tau_\quench$ turns out to be
  of the order of few times the inverse bandwidth. The decrease of
  $\Delta E(\rtime)$ for $\rtime>\tau_\quench$ can be fitted with a
  power law~(\ref{deltae-tau}) for $\rtime\gtrsim 10$.  The exponent
  turns out to be a rational number, which depends only on the phase in
  which the system is before the ramp (metallic phase for $U$ $<$ $2$,
  insulating phase for $U>2$) and after the ramp (metallic phase for
  $U$ $<$ $2V_f$, insulating phase for $U>2V_f$).  These results can
  be summarized as
  \begin{equation}
    \label{exe-fkm}
    \exE(\rtime) 
    \stackrel{\rtime\to\infty}{\sim}
    \left\{
      \begin{array}{ll}
        \rtime^{-{\textstyle\frac12}}
        & \text{linear ramp across the transition},
        \\[0mm]
        \rtime^{-1}
        & \text{linear ramp in metallic phase,}
        \\[0mm]
        \rtime^{-2}
        & \text{linear ramp in insulating phase.}
      \end{array}
    \right.
  \end{equation}

  How do these exponents arise and how do they depend on the ramp
  shape?  Further data show that the exponent $\eta$ $=$
  ${\textstyle\frac12}$ for the excitation across the metal-insulator
  transition is independent of the ramp shape $r(x)$. At present we
  have now simple explanation of this exponent. It would
  be interesting to determine how this exponent is related to the
  critical behavior of equilibrium correlation functions, such as the
  density of states at the transition~\cite{vanDongen1992a}.
  On the other hand, the behavior for ramps within either the metallic
  or the insulating phase will be explained in the next section by a
  perturbative argument, which applies to small ramps of arbitrary
  shape in any quantum system.  In particular we will see that the
  exponent $\eta$ $=$ $1$ is a consequence of the non-Fermi-liquid
  behavior of the metallic phase in the Falicov-Kimball model, while
  the exponent $\eta$ $=$ $2$ in the insulating phase is not an
  intrinsic property of the Falicov-Kimball model but is in fact due
  to the linear ramp shape.

  \section{Small ramps of arbitrary shape without traversing phase boundaries}
  \label{smallramp-sec-derivation}

  Our numerical results for ramps of the hopping amplitude in the
  Falicov-Kimball model show that the exponent $\eta$ in  Eq.~(\ref{exe-fkm}) 
  does not depend on the precise values of  the ramp parameters $V_i$ and
  $V_f$, but only on the thermodynamic phase of the initial and final state. 
  This finding suggests to study the excitation energy perturbatively
  in the limit of small ramp amplitudes, but for arbitrary ramp shapes
  and ramp durations.  In the remainder of this section we will derive
  the excitation energy $\Delta E(\rtime)$ up to second order in the
  ramp amplitude for an arbitrary Hamiltonian. In particular we  
  discuss the asymptotic behavior of $\Delta E(\rtime)$ in the limit
  $\rtime\to\infty$ and how it may be influenced by the ramp shape,
  and illustrate these general results with data for the specific case
  of the Falicov-Kimball model.

  \subsection{Perturbative result for the excitation energy}

  We consider the general Hamiltonian
  \begin{eqnarray}
    \label{adi-h}
    H(t) = H_0 + \kappa(t) W,
  \end{eqnarray}%
  where $H_0$ is the Hamiltonian before the ramp, $W$ is the operator
  that is switched on, and $\kappa(t)$ is the ramp function. As in
  Eq.~(\ref{vramp}), we characterize $\kappa(t)$ by the ramp amplitude
  $\ramp$, the ramp duration $\rtime$, and the ramp shape $r(x)$,
  i.e., $\kappa(t)$ $=$ $\ramp$ $r(t/\rtime)$.  In order to expand
  $\exE(\rtime)$ for fixed ramp duration $\rtime$ and ramp shape
  $r(x)$ in powers of $\ramp$, we decompose the quantum state
  $\ket{\psi(t)}$ of the system in the instantaneous eigenbasis
  $\ket{\phi_n(t)}$ of the Hamiltonian~(\ref{adi-h}), 
  which satisfies the condition
  \begin{eqnarray}
    \label{instanteigen}
    H(t)\ket{\phi_n(t)}
    =
    \epsilon_n(t)\ket{\phi_n(t)}
  \end{eqnarray}
  at any instance of time. We assume the $\ket{\phi_n(t)}$ to be
  nondegenerate for simplicity. After fixing the phase of the
  eigenvectors in a convenient way we obtain the eigenstate
  decomposition of $\ket{\psi(t)}$ as
  \begin{equation}
    \ket{\psi(t)} = \sum_n a_n(t) \,e^{i\int_0^t ds\, \epsilon_n(s)}\, \ket{\phi_n(t)},
  \end{equation}
  so that the Schr\"{o}dinger equation implies
  \begin{equation}
    \label{schroedinger-for-a}
    i\frac{d}{dt}a_n(t) = \sum_m 
    \,
    e^{i\int_0^t ds\, \epsilon_{nm}(s)}
    \,
    \big\langle\phi_n(t) \big|\frac{d}{dt}\big| \phi_m(t)\big\rangle,
  \end{equation}
  using the notation $\epsilon_{nm}(t)$ $=$ $\epsilon_{n}(t)-\epsilon_{m}(t)$. 
  The matrix element on the right-hand side of  Eq.~(\ref{schroedinger-for-a}) is given by
  \begin{eqnarray}
    \epsilon_{nm}(t)\,\big\langle\phi_n(t) \big|\frac{d}{dt}\big| \phi_m(t)\big\rangle  
    &=
    \big\langle\phi_n(t) \big|H(t)\frac{d}{dt}-\frac{d}{dt}H(t)\big| \phi_m(t)\big\rangle
    \nonumber\\
    &=
    \big\langle\phi_n(t) \big|\frac{dH}{dt}\big| \phi_m(t)\big\rangle 
    \nonumber\\
    \label{matrixelement}
    &=
    \ramp
    \,
    \frac{r'(t/\rtime)}{\rtime}
    \,
    \expval{\phi_n(t) |W| \phi_m(t)},
  \end{eqnarray}%
  where the first equality follows from Eq.~(\ref{instanteigen}) and
  the last from the explicit form of the Hamiltonian
  [Eq.~(\ref{adi-h})].  

  Because the system is assumed to be in the ground state
  $\ket{\phi_0(0)}$ of $H_0$ for $t$ $\le$ $ 0$, Eq.~(\ref{schroedinger-for-a})
  must be solved with the initial condition $a_m(0)=\delta_{m0}$. Together 
  Eq.~(\ref{matrixelement}) this implies that $a_n(t)$ $=$
  $\mathcal{O}(\ramp)$ for $n\neq 0$. In order to obtain the leading
  term in the expansion of $a_n(t)$ (for $n\neq 0$) we can thus
  restrict the sum in Eq.~(\ref{schroedinger-for-a}) to the single
  term $m$ $=$ $0$,
  \begin{equation}
    \label{ansowiesleibtundlebt}
    a_n(t) = \ramp \, \int_0^t 
    d\tb
    \;
    \frac{r'(\tb/\rtime)}{\rtime}
    \,
    \frac{\expval{\phi_n(\tb) |W| \phi_0(\tb)}}
    {\epsilon_{n0}(\tb)}
    \,
    e^{i\int_0^{\tb} ds \, \epsilon_{n0}(s)}
    \,+\,
    \mathcal{O}(\ramp^2).
  \end{equation}
  This expression was used before as a starting point for the
  discussion of ramps across a quantum critical point
  \cite{Polkovnikov2005a}. Here we study ramps which do not cross a
  phase boundary, and we assume that the instantaneous eigenenergies
  $\epsilon_{n0}(t)$ and eigenfunctions $\ket{\phi_n(t)}$, which
  depend on time $t$ only through the parameter $\kappa$, can be
  expanded around $\kappa=0$. Since $a_n(t)$ is already of order
  $\mathcal{O}(\ramp)$, $\epsilon_{n0}(t)$ and $\ket{\phi_n(t)}$ in
  Eq.~(\ref{ansowiesleibtundlebt}) can be replaced by $\epsilon_{n0}$
  $\equiv$ $\epsilon_{n0}(0)$ and $\ket{\phi_n}$ $\equiv$
  $\ket{\phi_n(0)}$, respectively.  The excitation energy, $\Delta
  E(\rtime)$ $=$ $\frac{1}{L}\sum_{n\neq 0} \epsilon_{n0}(t)$ $|a_n(t)|^2$, is
  then given by
  \begin{eqnarray}
    \label{fgr-res}
    \Delta E(\rtime) 
    &=&
    \ramp^2\,\mathcal{E}(\rtime) + \mathcal{O}(\ramp^3)
    \\
    \label{fgr-exe}
    \mathcal{E}(\rtime)
    &=&\,\,
    \int_0^\infty \frac{d\omega}{\omega}\,
    R(\omega) F(\omega\rtime) 
    \\
    \label{fgr-r}
    R(\omega)
    &=&\,\,
    \frac{1}{L}\sum_{n \neq 0} 
    \big|\expval{\phi_n |W| \phi_0}\big|^2
    \,\delta(\omega-\epsilon_{n0}) 
    \\
    \label{adiff}
    F(x) 
    &=&\,\,
    \left| \int_0^1  \!ds\,\, r'(s) e^{i x s} \right|^2.
  \end{eqnarray}%

  Eqs.~(\ref{fgr-res})-(\ref{adiff}) constitute the main result of
  this section. The correlation function $R(\omega)$, which can be
  interpreted as the spectral density of possible excitations induced
  by the operator $W$, is independent of the ramp shape $r(x)$ and the
  ramp duration $\rtime$. Conversely, the \emph{ramp spectrum} $F(x)$
  does not depend on the Hamiltonian but only on details of the ramp.
  For continuous ramp shapes $r(x)$ it follows that $F(x)$ $\to$ $0$
  for $|x|$ $\to$ $\pm \infty$, such that $F(\omega\rtime)$ becomes
  increasingly peaked around $\omega$ $=$ $0$ in the limit $\rtime$
  $\to$ $\infty$. In fact, making the replacement $F(\omega\rtime)$
  $\propto$ $\delta(\omega)/\rtime$ is equivalent to Fermi's Golden
  Rule for $|a_n(t)|^2$, and the nonadiabatic
  excitation~(\ref{fgr-exe}) is due to deviations of $F(\omega\rtime)$
  from $\delta(\omega)$.
  The crossover scale $\tau_\quench$ that was discussed in
  Sec.~\ref{secfiniteramps} is thus given by the value of $\rtime$ below which 
  $F(\omega\rtime)$ $=$ $F(0) + \mathcal{O}(\omega^2\rtime^2)$ is approximately constant
  over the entire bandwidth $\Omega$ of $R(\omega)$, i.e., $\tau_\quench$ $\approx$ $1/\Omega$.
  In  Sec.~\ref{subsec:metallic} we confirm this estimate numerically
  for small interaction ramps in the metallic phase of the
  Falicov-Kimball model.

  In the following we will analyze the asymptotic behavior of
  Eq.~(\ref{fgr-exe}) in the adiabatic limit, $\rtime$ $\to$ $\infty$.
  For this we need the behavior of the ramp spectrum $F(x)$ at large
  values of $x$, which follows from Eq.~(\ref{adiff}) as
  \begin{eqnarray}
    \label{asy-f}
    F(x) \stackrel{x\to\infty}{\sim} \frac{f(x)}{x^{\alpha}},
    \text{~~with~~}
    f(x) \stackrel{x\to\infty}{=} \mathcal{O}(1),    
  \end{eqnarray}
  where the exponent is given by $\alpha=2n$ if the $n$th derivative
  of $r(x)$ is discontinuous (i.e., the ($n-1$)st derivative has a
  kink), but all lower derivatives are continuous; this behavior
  follows from the Riemann-Lebesgue lemma\cite{Olver1997a}.  For
  example, in case of a linear $r(x)=x$ the first
  derivative $r'(x)=\Theta(x)\Theta(1-x)$ is discontinuous at $x=0$
  and $x=1$, so that the ramp spectrum $F(x)$ decays like $x^{-2}$
  [cf.  Eq.~(\ref{rampr2-f}) below]. In general, when the ramp shape 
  has a finite number of kinks, the large-$x$ asymptotics of the
  ramp spectrum (\ref{adiff}) is given by a finite sum of oscillating 
  terms, $F(x)\sim|\sum_k f_k \cos(\omega_k x)+\delta_k|^2/x^\alpha$. 
  By choosing a smooth ramp one can always increase the 
  exponent $\alpha$ or even make $F(x)$ decay exponentially for $x\to\infty$. 
  However, in practice ramp protocols often have kinks that lead to a 
  power-law decay~(\ref{asy-f}).

  To estimate the magnitude of the integral~(\ref{fgr-exe}) in the
  limit $\rtime$ $\to$ $\infty$ we distinguish two cases, namely (i)
  the \emph{gapless} case, in which $R(\omega)$ vanishes like a power
  law at $\omega=0$, and (ii), the case of a \emph{gapped} excitation
  spectrum, in which $R(\omega)$ has a finite gap
  $\Omega_{\text{gap}}$ above $\omega$ $=$ $0$.  In both cases we
  assume that $R(\omega)$ is zero beyond some high-frequency scale
  $\Omega$, although the argument remains valid if $R(\omega)$
  vanishes exponentially for $\omega>\Omega$. 
  Since $\int_0^{\infty}d\omega\;R(\omega)$ $=$ $\frac{1}{L}\big[\bra{\phi_0}W^2\ket{\phi_0}$ $-$ $\bra{\phi_0}W\ket{\phi_0}^2\big]$ 
  we can also assume that any singularities of $R(\omega)$ are integrable.

  \subsection{Case (i): Gapless excitation spectrum}
  
  In this paragraph we discuss the case in which the excitation spectrum $R(\omega)$
  is gapless and vanishes like a power law at $\omega=0$,
  \begin{equation}
    \label{r-casei}
    R(\omega) \stackrel{\omega\to 0}{\sim} \omega^\nu
    \text{~~with~~}
    \nu>0.
  \end{equation}
   First we assume $\alpha>\nu$, where $\alpha$ is the exponent that characterizes 
  the ramp shape [Eq.~(\ref{asy-f})].  Writing $R(\omega)=\omega^\nu \,\tilde R(\omega)$, 
  the integral~(\ref{fgr-exe}) becomes, after a change of variables,
  \begin{equation}
    \label{exe-case1}
    \mathcal{E}(\rtime)
    =
    \frac{1}{\rtime^\nu}
    \,
    \int_0^{\Omega\rtime} 
    \!dx\, 
    x^{\nu-1}\, 
    \tilde R(x\rtime^{-1})
    F(x).
  \end{equation}
  Using the asympotic  behavior~(\ref{asy-f}) we find that the integral
  in this expression remains finite in the limit $\rtime\to\infty$, so that in this case
  the leading contribution to the excitation energy is given by
  \begin{eqnarray}
    \label{exe-case11}
    \alpha>\nu\text{:~~~~~~}&
    \mathcal{E}(\rtime)
    \stackrel{\rtime\to\infty}{\sim}
    \frac{C}{\rtime^\nu}
    \equiv
    \mathcal{E}_{\text{intr}}(\rtime),
  \end{eqnarray}
  with $C$ $=$ $\tilde R(0) \int_0^{\infty} dx\,x^{\nu-1} F(x)$.  It
  is important to note that the exponent does not depend on the ramp
  shape, but only on the density of possible excitations above
  $\omega=0$. Because the latter is an intrinsic property of the
  system we will refer to $\mathcal{E}_{\text{intr}}(\rtime)$ as the
  \emph{intrinsic} contribution to the excitation energy in the
  following. In principle, a ramp between two parameter values can
  always be made so smooth that  $\mathcal{E}_{\text{intr}}(\rtime)$ becomes 
  the dominating contribution to the excitation energy (i.e., $\alpha>\nu$), 
  as in Eq.~(\ref{exe-case11}). However, as we will see in the following 
  paragraph, if the ramp is not smooth enough (i.e., if $\alpha\le\nu$), the intrinsic
  contribution will be masked by a nonuniversal contribution
  that is essentially determined by the ramp shape.

   For the case of $\alpha\le\nu$ we estimate the
   integral~(\ref{fgr-exe}) as follows. For the moment we assume that 
   the spectral density  $R(\omega)$ has no singularities at finite frequencies and
   use the bounds
  \begin{equation}
    \label{bound-r}
    \omega^\nu 
    \,C_1\,
    \Theta(\Omega_1-\omega)
    \le 
    R(\omega)
    \le
    \omega^\nu
    \,C_2\,
    \Theta(\Omega_2-\omega),
  \end{equation}
  with positive constants $C_1$, $C_2$, $\Omega_1$, $\Omega_2$.
  Together with Eqs.~(\ref{asy-f}) and~(\ref{fgr-exe}) we obtain for
  the excitation energy for $\rtime\to\infty$,
  \begin{eqnarray}
    \label{de-case12}
    \alpha<\nu\text{:~~~~~~}&
    \frac{C_1'}{\rtime^{\alpha}}
    % &
    \,\le\,
    \mathcal{E}(\rtime)
    \,\le\,
    % &
    \frac{C_2'}{\rtime^{\alpha}} 
    % \text{~~~~~~~~~~~~}
    \\
    \label{de-case13}
    \alpha=\nu\text{:~~~~~~}&
    C_1'\frac{\,\log(\rtime\Omega)}{\rtime^{\alpha}} 
    % &
    \,\le\,
    \mathcal{E}(\rtime)
    \,\le\,
    % &
    C_2'\frac{\,\log(\rtime\Omega)}{\rtime^{\alpha}},
    % \text{~~~}
    % &\alpha = \nu
  \end{eqnarray}
  with positive constants $C_1'$ and $C_2'$.  
  The upper bound holds because $f(x)=\mathcal{O}(1)$ 
  in Eq.~(\ref{asy-f}). To obtain the lower bound it is sufficient to 
  note that although $f(x)$ can have infinitely many zeros, the moving 
  average  $\bar f(x) = \int_x^{x+h} dx f(x)$  over any small finite interval
  of given length $h$ is larger than some positive constant. This property is 
  satisfied in particular when the ramp shape has a finite number of kinks, 
  as discussed below Eq.~(\ref{asy-f}).
  Finally we note that Eqs.~(\ref{de-case12}) and (\ref{de-case13})
  hold also if $R(\omega)$ has integrable singularities, because a
  small frequency interval around each of them contributes to the
  integral~(\ref{fgr-exe}) in the same way as the gapped spectrum
  [Sec.~\ref{subsec:gapped}], namely $\propto$ $\tau^{-\alpha}$
  [Eq.~(\ref{exe-case2})].

  The result that is stated in Eqs.~(\ref{de-case12}) and (\ref{de-case13}) has 
  a simple  interpretation: Kinks in the ramp shape increase the probability of
  excitations to high energy states, as expressed by the slowly decaying
  tail of the ramp spectrum $F(x)$. When the ramp is not smooth enough
  the integral~(\ref{fgr-exe}) is therefore dominated by the
  high-frequency part of $R(\omega)$, leading to a nonuniversal,
  ramp-shape dependent excitation energy. For the often-considered
  linear ramp ($\alpha$ $=$ $2$) any intrinsic contribution
  $\mathcal{E}_{\text{intr}}(\rtime)$ with $\nu$ $\ge$ $2$ will
  therefore be unobservable in $\mathcal{E}(\rtime)$. This is
  precisely what happens for weak-coupling interaction ramps in the
  Hubbard model, as discussed below in Sec.~\ref{subsec:metallic}.

  \subsection{Case (ii): Gapped excitation spectrum}
  \label{subsec:gapped}

  We now turn to the case of an excitation density which has a gap
  $\Omega_{\text{gap}}$ at $\omega=0$. The integral~(\ref{fgr-exe})
  then starts at the finite lower bound $\Omega_{\text{gap}}$, such
  that $F(x)$ can be replaced by its asymptotic behavior~(\ref{asy-f})
  in the entire integration range. As a consequence we have
  \begin{equation}
    \label{exe-case2}
    \mathcal{E}(\rtime)
    \stackrel{\rtime\to\infty}{\sim}
    \frac{1}{\rtime^\alpha}
    \,\,
    \int_{\Omega_{\text{gap}}}^{\Omega}
    \!\!dx\,\,
    \frac{R(x) f(x\rtime)}{x^{\alpha+1}}.
  \end{equation}
  The integral gives a finite constant in the limit $\rtime\to\infty$
  provided that $R(\omega)$ is not singular. Otherwise the integral
  may give a $\rtime$-dependent but bounded contribution, as shown in
  the next subsection for ramps in the insulating phase 
  of the Falicov-Kimball model.
  The gapped case [Eq.~(\ref{exe-case2})] is thus similar to the
  gapless case with $\alpha$ $<$ $\nu$ [Eq.~(\ref{de-case12})].  In
  both cases the excitation energy is dominated the high-frequency
  behavior of $F(x)$ and is therefore completely determined by the
  ramp shape, while the intrinsic contribution~(\ref{exe-case11}) is
  unobservable.

  Our analysis so far can be summarized as follows. The excitation
  energy after a ramp may be either dominated by the intrinsic
  contribution~(\ref{exe-case11}) or set by ramp-shape dependent terms
  [Eqs.~(\ref{de-case12}), (\ref{de-case13}), and~(\ref{exe-case2})],
  depending on the large-frequency asymptotics~(\ref{asy-f}) of the
  ramp spectrum and the small-frequency behavior~(\ref{r-casei}) of
  the excitation density.  This fact will be illustrated in the
  following two subsections for ramps in the insulating and metallic
  phase of the Falicov-Kimball model and the Hubbard model.

  \begin{figure}
    \centerline{\includegraphics[width=0.9\columnwidth]{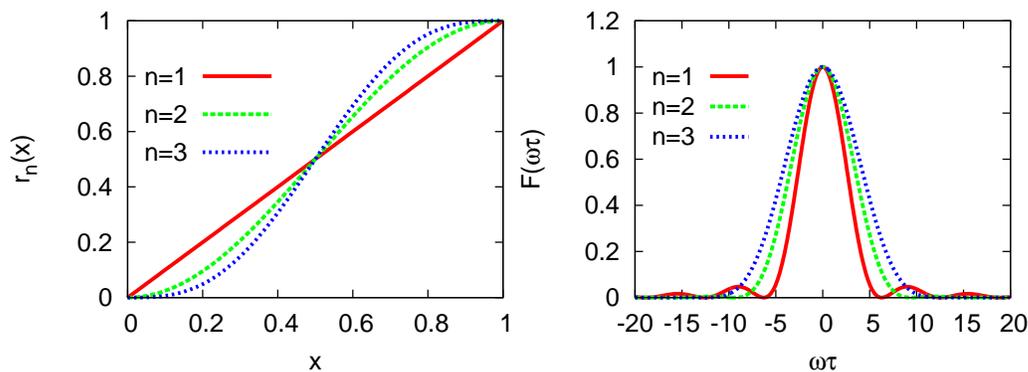}}
    \caption{
      Ramps shapes given by Eqs.~(\ref{rampr1})-(\ref{rampr3}) (left panel),
      and corresponding ramp spectra [Eqs.~(\ref{rampr1-f})-(\ref{rampr3-f})]
      (right panel). The Fresnel oscillations in $F_n(\omega)$ are due to the 
      discontinuity in the derivatives of $r_n(x)$ at $x$ $=$ $0$ and $x=1$.
    }
    \label{fig-ramps}
  \end{figure}

  \subsection{Insulating phase}
  \label{subsec:insulating}

  In previous subsection we have shown that the excitation energy after a 
  ramp within a gapped phase behaves in a nonuniversal way because the 
  intrinsic contribution~(\ref{exe-case11}) vanishes. A significant dependence 
  of the nonadiabatic excitation energy on the ramp shape is therefore expected 
  also for ramps with finite amplitude. In the following we will demonstrate this 
  fact for ramps within the insulating phase of the Falicov-Kimball
  model, where it turns out that the asymptotic behavior for $\rtime\to\infty$
  is indeed correctly described by the analytic expression~(\ref{exe-case2}) 
  that was obtained for small ramps.

  For this purpose we focus on three particular ramp shapes,
  \numparts%
  %\label{nonliner-ramps}
  \begin{eqnarray}%
    \label{rampr1}
    r_1(x) &= x
    \\
    \label{rampr2}
    r_2(x) &= \frac{1-\cos(\pi x)}{2} 
    \\
    \label{rampr3}
    r_3(x) &= \frac{\pi x-\cos(\pi x)\sin(\pi x)}{\pi}.
  \end{eqnarray}%
  \endnumparts%
  Here $r_n(x)$ is chosen in such a way that its $n$th derivative is
  discontinuous at $x=0$ and $x=1$ (Fig.~\ref{fig-ramps}a), i.e., $r_n'(x)$
  $\propto$ $\sin^n(\pi x)$ for $0 < x < 1$. The corresponding 
  ramp spectra [Eq.~(\ref{adiff})] are
  \numparts%
  %\label{nonliner-ramps-f}
  \begin{eqnarray}
    \label{rampr1-f}
    F_1(\omega) &= 2\,\frac{1-\cos(\omega)}{\omega^2}
    \\
    \label{rampr2-f}
    F_2(x) &=\frac{\pi^4}{2}\frac{1+\cos(\omega)^2}{(\pi^2-\omega^2)^2}
    \\
    \label{rampr3-f}
    F_3(x) &= 32 \pi^4\frac{1-\cos(\omega)}{\omega^2(4\pi^2-\omega^2)^2}.
  \end{eqnarray}%
  \endnumparts%
  These functions vanish like $F_n(x)$ $\sim$ $x^{-2n}$ for
  $x\to\infty$ (Fig.~\ref{fig-ramps}b).  We now perform ramps of the
  hopping amplitude $V(t)=V_i + (V_f-V_i)\,r_n (t/\rtime)$ in the
  Falicov-Kimball model, with $V_i$ $=$ $V$ $=$ $1$ as the energy
  scale. We consider only the paramagnetic insulating phase at
  half-filling, i.e., $U>2=2V_i$ and $U>2V_f$.

\begin{figure}
    \centerline{\includegraphics[width=0.85\columnwidth]{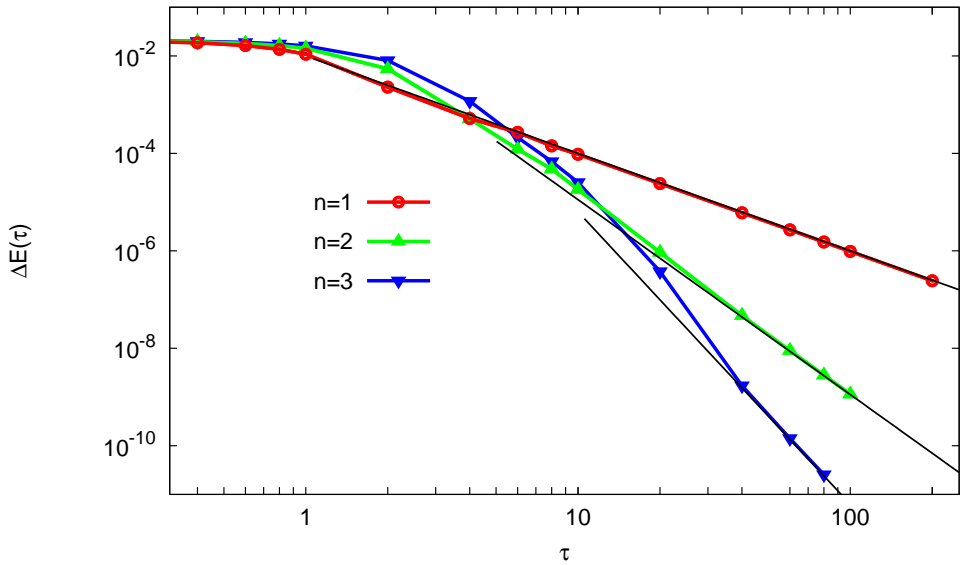}}
    \caption{
      Same as Fig.~\ref{fig-exe1}, for ramps within the insulating phase.
      $r(x)=r_n(x)$ [Eqs.~(\ref{rampr1})-(\ref{rampr3})], $U=3$, $V_f=0.5$. The
      black linear lines correspond to power-law behavior~(\ref{deltae-tau})
      with $\eta=2$ ($n=1$) and $\eta=4$ ($n=2$), and  $\eta=6$ ($n=3$).
    }
    \label{fig-adiins}
  \end{figure}

  The excitation energy after such ramps is plotted as a function of
  the ramp duration $\rtime$ in Fig.~\ref{fig-adiins}. The curves can
  be fitted with power laws~(\ref{deltae-tau}) for large $\rtime$,
  with an exponent $\eta$ $=$ $2$ for the linear ramp~(\ref{rampr1})
  and $\eta$ $=$ $4$ for the ramp~(\ref{rampr2}), respectively. For
  the ramp~(\ref{rampr3}) the results are consistent with an exponent
  $\eta$ $=$ $6$, but the excitation energy is too small for a
  power-law fit in the accessible range. Hence the large-$\rtime$
  behavior of the nonadiabatic excitation in the case of ramps with
  finite amplitude turns out to be the same as in the limit of small
  ramps, i.e., a power law with an exponent that is determined by the
  singularities of the derivatives of the ramp shape
  [cf.~Eq.~(\ref{exe-case2})] rather than by intrinsic properties of
  the system.

\begin{figure}[t]
    \centerline{\includegraphics[width=0.85\columnwidth]{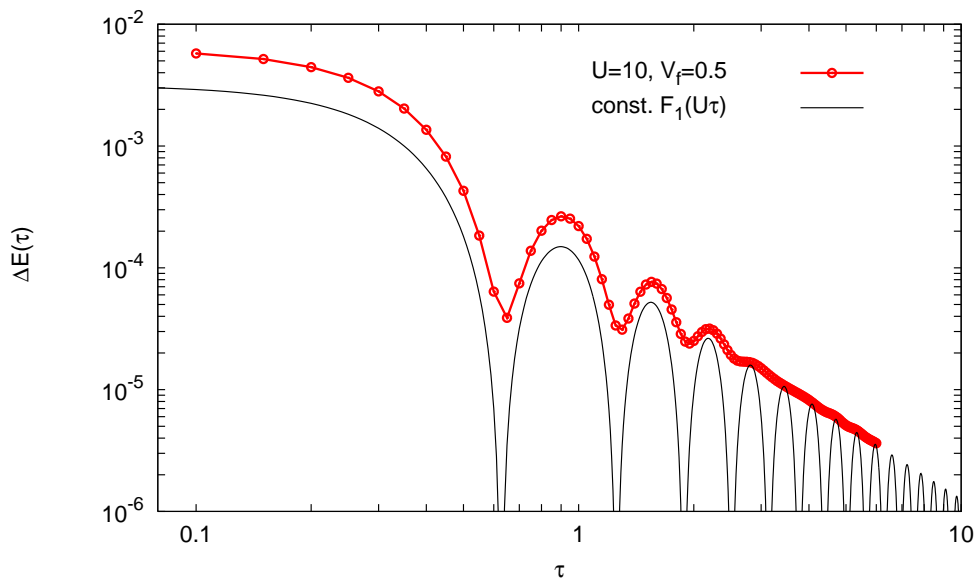}}
    \caption{
      Excitation energy \ref{exe} after the linear ramp~(\ref{rampr1}) 
      of the hopping amplitude within the insulating phase [$U=10$, $V_i=1$, $V_f=0.5$],
      compared to Eq.~(\ref{exe-alimit}) for small ramp amplitudes
      [$F_1$ given by Eq.~(\ref{rampr1-f})].
    }
    \label{fig-rampins1}
  \end{figure}

  To check Eq.~(\ref{fgr-exe}) explicitly for arbitrary ramps we would
  have to compute the density of excitations $R(\omega)$, for which no
  general solution is available. Nevertheless one can derive an
  expression in the atomic limit and compare the resulting excitation
  energy to ramps deep in the insulating phase. For ramps of the
  hopping $V(t)$, the operator $W$ in Eq.~(\ref{fgr-r}) is given by the
  kinetic energy operator. In the case of half-filling for both mobile
  and immobile particles there is exactly one particle per site in the
  ground state for $V=0$. Therefore each hopping process creates
  exactly one doubly-occupied site, and the function $R(\omega)$
  consists of a single delta peak at $\omega=U$. When $R(\omega)$
  $\propto$ $\delta(\omega-U)$ is inserted into Eq.~(\ref{fgr-exe})
  one obtains
  \begin{equation}
    \label{exe-alimit}
    \exE(\rtime)\propto F(U\rtime),
  \end{equation}
  i.e., the Fresnel oscillations in ramp spectra such as  Eqs.~(\ref{rampr1-f}) 
  to~(\ref{rampr1-f}) become visible in the dependence of the excitation 
  energy on $\rtime$. This result should only be slightly modified for ramps 
  deep in the insulating phase ($U\gg V$), assuming that the delta-peak is 
  then only slightly broadened and shifted in position.  In fact, as seen in
  Fig.~\ref{fig-rampins1} the tail oscillations of $F_1(x)$
  [Eq.~(\ref{rampr1-f}), Fig.~\ref{fig-ramps}] are also apparent in the excitation 
  energy for ramps between states with $U$ $\gg$ $V$.  In
  the limit $\rtime$ $\to$ $\infty$ these oscillations are washed out
  because $R(\omega)$ has a finite bandwidth $\Delta\Omega$ for $V$ $>$ $0$,
  such that the integral~(\ref{fgr-exe}) averages over many oscillations 
  for $\rtime\gg1/\Delta\Omega$.  

  \subsection{Metallic phase}
  \label{subsec:metallic}

  As an application of Eq.~(\ref{fgr-exe}) to ramps in a gapless 
  phase we consider the turn-on of the interaction in the Falicov-Kimball 
  model and the Hubbard model. For the following discussion it is convenient
  to write the Hamiltonian in momentum space
  \begin{eqnarray}%
    \label{hubbard}
    H 
    &= \sum_{\Vk \sigma} (\epsilon_{\Vk\sigma}-\mu_\sigma) c_{\Vk\sigma}^{\mydagger} c_{\Vk\sigma}^{\phdagger}
    + U(t) D 
    \\
    \label{wint}
    D
    &=
    \sum_i n_{i\downarrow}n_{i\uparrow}
    =
    \sum_{\Vk,\Vk',\Vq}
    c_{\Vk+\Vq \downarrow}^\mydagger c_{\Vk \downarrow}^{\phdagger}
    c_{\Vk'-\Vq \uparrow}^\mydagger c_{\Vk' \uparrow}^{\phdagger}.
  \end{eqnarray}%
  Furthermore we change the notation with respect to Eq.~(\ref{FKM}) to allow 
  for a unified description of the Hubbard model, where both spin species are 
  mobile ($\epsilon_{\Vk\uparrow}$ $=$ $\epsilon_{\Vk\downarrow}$, $\mu_\uparrow$ $=$ $\mu_\downarrow$) 
  and the Falicov-Kimball model, where we take spin $\uparrow$ to be immobile 
  ($\epsilon_{\Vk\uparrow}$ $=$ $0$, $\mu_\uparrow$ $=$ $\mu-E_f$, $\mu_\downarrow$ $=$ $\mu$). We consider ramps 
  at half filling ($\mu_\sigma$ $=$ $U/2$) in which the interaction is changed from 
  zero to a finite value, $U(t)=\Delta U r(t/\rtime)$.

  For a ramp of the interaction strength in the Hubbard model and the 
  Falicov-Kimball model, the operator $W$ in Eq.~(\ref{adi-h}) is given by 
  the double occupation~(\ref{wint}). As shown in \ref{app2}, the 
  excitation density $R(\omega)$ at $U=0$ can be expressed in terms of the 
  second-order contribution to the self-energy
  \begin{equation}
    \label{rrealfrequencymain}
    R(\omega) =
    -\frac{1}{\pi}
    \sum_{-\omega \le \epsilon_{\Vq\downarrow}\le 0} \,
    \text{Im}\;\Sigma^{(2)}_{\Vq\downarrow}(\omega+\epsilon_{\Vq\downarrow}+i0).
  \end{equation}
  For comparison to our DMFT results we evaluate
  Eq.~(\ref{rrealfrequencymain}) in the limit of infinite dimensions
  \cite{Metzner1989a} where the self-energy is independent of momentum
  $\Vq$, \cite{Muellerhartmann1989a} and the $\Vq$-summation can be
  replaced by an integral over the density of states
  $\rho_{\downarrow}(\epsilon)$,
  \begin{equation}
    \label{r-dmft-u0}
    R(\omega) = -\frac{1}{\pi}
    \int_{-\omega}^0\! d\epsilon\, \rho_\downarrow(\epsilon)
    \,\text{Im}\;\Sigma^{(2)}_\downarrow(\epsilon+\omega+i0).
  \end{equation}
  For the Hubbard model, the second-order self-energy is 
  given by \cite{Muellerhartmann1989a}
  \begin{equation}
    \label{sigma2-hub}
    -\frac{1}{\pi}\text{Im}\; \Sigma^{(2)}_{\text{Hub}}(\omega) = 
    \int_0^\omega
    \!\!d\mu\,
    \rho(\mu-\omega)
    \int_0^\mu 
    \!\!d\nu\,
    \rho(\nu)\rho(\mu-\nu)
    \stackrel{\omega\to0}{\sim}
    {\textstyle\frac{1}{2}}\,{\rho(0)^3}\,\omega^2.
  \end{equation}
  In accordance with Fermi liquid theory the imaginary part of the
  self-energy vanishes $\propto \omega^2$, thus leading to
  well-defined quasiparticle excitations in the metallic phase of the
  Hubbard model.  On the other hand, the imaginary part of the mobile
  electron self-energy in the Falicov-Kimball model remains finite at
  $\omega=0$ due to the scattering off fixed impurities. Its value can
  be obtained easily from the exact solution of the Falicov-Kimball
  model in DMFT~\cite{Freericks2003a},
  \begin{eqnarray}
    \label{sigma2-fkm}
    -\frac{1}{\pi}\text{Im}\; \Sigma^{(2)}_{\text{FKM}}(\omega)
    = (1-n_f)n_f \rho(\omega)
    \stackrel{\omega\to0}{\sim} (1-n_f)n_f \rho(0).
  \end{eqnarray}
  Here $n_f$ is the average density of localized particles, such that
  $n_f=0.5$ in case of half-filling.  Eqs.~(\ref{sigma2-hub})
  and~(\ref{sigma2-fkm}) can then be inserted in
  Eq.~(\ref{r-dmft-u0}), which in turn determined the intrinsic
  component~(\ref{exe-case11}) of the excitation energy,
  \begin{eqnarray}
    \label{adi-met-hub}
    \text{Hubbard:~~}  
    &R(\omega) 
    \stackrel{\omega\to0}{\sim}
    {\textstyle\frac{1}{6}}\,\rho(0)^4\,\omega^3 
    &\text{~}\Rightarrow\text{~}
    \mathcal{E}_{\text{intr}}(\rtime) \propto \rtime^{-3}
    \\
    \label{adi-met-fkm}
    \text{Falicov-Kimball:~~}  
    &R(\omega) 
    \stackrel{\omega\to0}{\sim}
    (1-n_f)n_f \rho(0)^2\,\omega 
    &\text{~}\Rightarrow\text{~}
    \mathcal{E}_{\text{intr}}(\rtime) \propto \rtime^{-1}.
  \end{eqnarray}%

  As discussed above, the intrinsic contribution can be masked
  completely by a ramp-shape dependent contribution if the ramp is not
  smooth enough, i.e., when the exponent in Eq.~(\ref{asy-f})
  satisfies $\alpha \le 1$ or $\alpha \le 3$ in case of the
  Falicov-Kimball and Hubbard model, respectively. However, the
  discussion below Eq.~(\ref{asy-f}) shows that $F(x)$ decays at least
  $\propto x^{-2}$ if the ramp is continuous, i.e., if it does not
  contain any abrupt finite changes. Hence we conclude that the
  intrinsic component~(\ref{adi-met-fkm}) is always dominant for ramps
  that turn on a small interaction in the Falicov-Kimball model.  This
  is consistent with our numerical results for ramps in the metallic
  phase of the Falicov-Kimball model which are not shown here, namely,
  that the $1/\rtime$ behavior for the metallic phase
  in~(\ref{exe-fkm}) does not only hold for linear ramps
  (Fig.~\ref{fig-exe1}), but for all three ramps~(\ref{rampr1})
  to~(\ref{rampr3}).

  The situation is very different for ramps in the Hubbard model.
  Because the intrinsic contribution~(\ref{adi-met-hub}) vanishes
  $\propto\rtime^{-3}$ for $\rtime\to\infty$ it is negligible with
  respect to the high-frequency contribution~(\ref{de-case12}) for
  linear ramps, where the ramp spectrum decays as $x^{-2}$
  [Eq.~(\ref{rampr1-f})]. This is consistent with results of M\"ockel
  and Kehrein \cite{Moeckel2009b}, who computed the excitation energy
  after a linear ramp of the interaction in the Hubbard model using
  Keldysh perturbation theory and found $\Delta E(\rtime) \sim
  \rtime^{-2}$ for $\rtime\to\infty$.

\begin{figure}
    \centerline{\includegraphics[width=0.85\columnwidth]{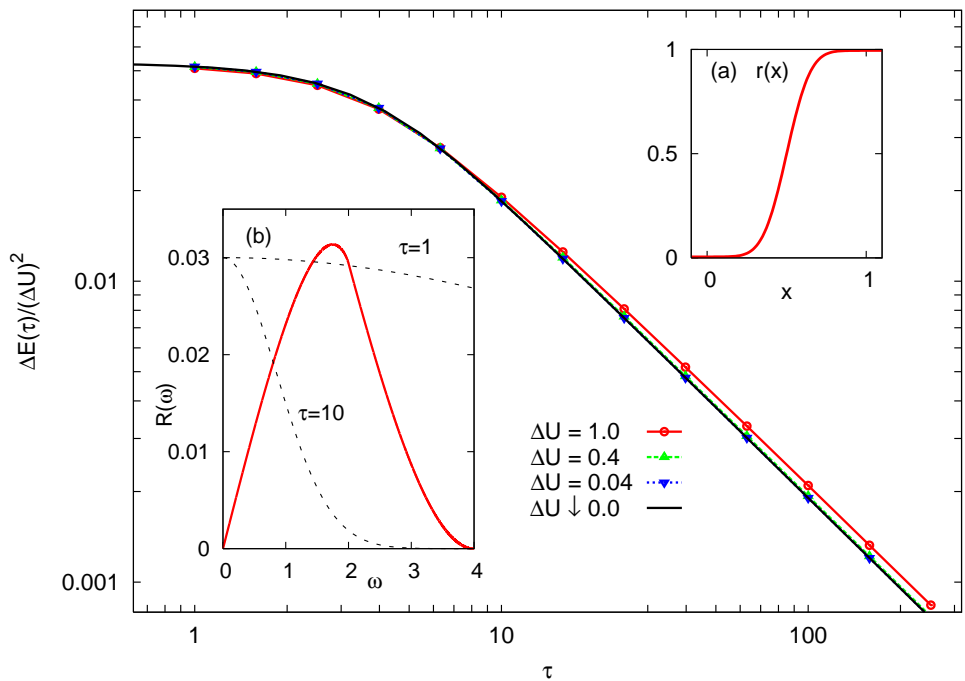}}
    \caption{
      Excitation $\Delta E(\rtime)$ in metallic phase of the
      Falicov-Kimball model, after Gaussian ramps
      [Eq.~(\ref{gaussramp-ramp}), $c_2=36$, see inset (a)] 
      from interaction $U$ $=$ $0$ to $\Delta U$. The curve $\Delta U=0$ 
      is obtained from  Eq.~(\ref{fgr-exe}), where $R(\omega)$ is obtained 
      from  Eqs.~(\ref{r-dmft-u0}) and~(\ref{sigma2-fkm}), and $F(\omega)$
      is given by Eq.~(\ref{gaussramp-ramp}). The excitation spectrum
      $R(\omega)$ (solid line) is plotted in the inset (b), together with
      $0.03\,F(\omega\rtime)$ (dashed lines) to illustrate the effect
      of $\rtime$ (see text).
    }
    \label{fig-ramp-met}
  \end{figure}

  The exact expression for $R(\omega)$ at all frequencies in the
  Falicov-Kimball model at $U=0$ [Eqs.~(\ref{r-dmft-u0})
  and~(\ref{sigma2-fkm})] allows us to evaluate $\mathcal{E}(\rtime)$
  for arbitrary interaction ramps $U(t)=\Delta U r(t\rtime)$ at finite
  ramp durations, and compare to the DMFT result for $\Delta E(\rtime)$.
  In Fig.~\ref{fig-ramp-met}
  this comparison is done for a smooth
  \emph{Gaussian ramp} [inset (a) in Fig.~\ref{fig-ramp-met}], which we define by
  \begin{eqnarray}
    \label{gaussramp-ramp}
    r'(x) &= c_1 \exp\big[-c_2 (x-1/2)^2\big], 
    \\
    \label{gaussramp-ramp-f}
    F(x) &=\frac{\pi c_1^2}{c_2} \exp\Big(-\frac{x^2}{4c_2}\Big)
    \text{~~for~~$c_2\gg1$},
  \end{eqnarray}%
  where $c_1$ is a normalization constant to satisfy the condition
  $\int_0^1 \!dx \,r'(x)$ $=$ $r(1) -r(0)$ $=$ $1$, and $c_2$ is
  chosen such that $r'(x)$ is sufficiently small at the boundary $x=0$
  and $x=1$, i.e., the expression for $F(x)$ holds up to terms which
  are exponentially small in $c_2$. For $U\lesssim 1$ numerical
  results for the excitation energy (scaled with the ramp amplitude
  $\Delta U^2$) agree very well with the analytical
  expression~(\ref{fgr-exe}), evaluated using Eqs.~(\ref{r-dmft-u0}),
  (\ref{sigma2-fkm}), and~(\ref{gaussramp-ramp-f}). This corroborates
  the validity argument of Sec.~\ref{smallramp-sec-derivation} and
  shows that it provides correct estimates for the nonadiabatic
  excitation energy after ramps which are not too large in amplitude.

  Inset (b) of Fig.~\ref{fig-ramp-met} illustrates the origin of the
  crossover from small to large ramp times $\rtime$. For fast ramps,
  e.g., $\rtime=1$, the ramp spectrum $F(\omega\rtime)$ averages over
  the entire bandwidth $\Omega$ of the excitation spectrum
  $R(\omega)$. As a consequence, the excitation energy becomes
  independent for ramp times smaller than the quench time scale
  $\rtime\lesssim1/\Omega=\tau_\quench$.  Indeed we see in the
  numerical data that the weak-coupling quench time scale fits with
  the estimate $1/\Omega\approx1/4$. For larger quench times,
  e.g., $\rtime\gtrsim10$ in inset (b) of Fig.~\ref{fig-ramp-met}, 
  the ramp spectrum $F(\omega\rtime)$ probes only the linear small-$\omega$ 
  behavior of $R(\omega)$, leading to the universal power-law
  behavior $\Delta E(\rtime)\sim1/\rtime$ for $\rtime\gtrsim10$.

  \section{Conclusion}
  \label{concl}

  We presented a general perturbative analysis of the excitation
  energy due to slow ramps of a parameter in a quantum system without
  crossing of phase boundaries, motivated by our numerical results for
  the Falicov-Kimball model obtained with nonequilibrium dynamical
  mean-field theory. We demonstrated that the excitation energy
  vanishes algebraically for large ramp duration $\tau$ [Eq.~(\ref{deltae-tau})] 
  under rather general circumstances. The exponent $\eta$ can depend on 
  one hand on the spectrum of the correlation function of the operator 
  that is switched on, and on the other hand on the differentiability 
  of the ramp function. Which of these influences dominates in $\eta$ 
  depends on the low-energy  behavior of the excitation spectrum compared 
  to the spectrum of the ramp protocol. In practice, any experimental ramp 
  protocol can always be considered as differentiable on a short enough 
  timescale. Our conditions on the degree of differentiability have to be 
  interpreted in the sense that a ramp protocol must be considered as not 
  differentiable if the slope or any higher derivative changes on a timescale 
  shorter than the inverse bandwidth of the system.
  
  For ramps in gapped systems the asymptotic behavior of the
  excitation energy depends only on the ramp protocol and can be made
  as small as desired by use of increasingly smooth ramp shapes.  By
  contrast, for ramps in gapless systems the low-energy excitation
  spectrum has no effect on $\eta$ if the ramp is not smooth enough.
  Only if the ramp is sufficiently smooth does $\eta$ become
  ramp-independent and reflects the low-energy excitation spectrum of
  the system.  For the fermionic Hubbard model this implies that a
  linear ramp from $U$ $=$ 0 to a small value of $U$ leads to an
  unnecessarily large excitation energy with $\eta$ $=$ 2, which can be
  reduced to the intrinsic exponent $\eta$ $=$ 3 if the ramp shape has
  at least two continuous derivatives.

  Our results also indicate that in the Falicov-Kimball model the
  exact expression for the excitation energy in the limit of small
  ramps provides a good estimate up to quite large ramp amplitudes.
  This suggests to use the perturbative expression, which is valid for
  arbitrary systems, as a guide for finding ramp protocols 
  that connect fixed parameters of the Hamiltonian and minimize 
  the excitation energy for a given ramp time, thereby improving 
  the preparation of states in experiments with ultracold atomic 
  gases.

  \section*{Acknowledgments}

  Useful discussions with Stefan Kehrein are gratefully acknowledged.
  M.E.\ acknowledges support by Studienstiftung des deutschen Volkes.
  This work was supported in part by SFB 484 of the Deutsche
  Forschungsgemeinschaft.

  \appendix
  
  \section{Solution of the Falicov-Kimball model in nonequilibrium
    using DMFT\protect\footnote{In the appendices, $\tau$ denotes
      imaginary time, not to be confused with the ramp duration $\tau$
      in the main text.}}
  \label{app1}

  In this appendix we describe in some detail how the Falicov-Kimball
  model~(\ref{FKM}) with arbitrary time-dependent hopping amplitude
  $V(t)$ or time-dependent interaction $U(t)$ is solved using
  nonequilibrium DMFT. In DMFT, local correlation functions of the
  lattice model are obtained from an effective impurity model in which
  a single site is coupled to a self-consistently determined
  environment~\cite{Georges1996a}. The mapping of the lattice model
  onto the single-site model, which becomes exact in the limit of
  infinite dimensions~\cite{Metzner1989a}, can be formulated either in
  imaginary time, which yields a theory for thermal equilibrium, or in
  real time (using the Keldysh technique), which yields a theory that
  can be applied to various nonequilibrium
  situations~\cite{Schmidt2002a}.
  For the Falicov-Kimball model the action of the single-site model
  can be reduced to a quadratic one~\cite{Brandt1989a}, such that
  nonequilibrium correlations function can be determined from a closed
  set of equations of 
  motion~\cite{Freericks2006a,Freericks2008b,Eckstein2008a,Tran2008a,Joura2008a,Tsuji2008a,Tsuji2009a}.

  Because we are interested in the transient time evolution of a
  system which is in thermal equilibrium for times $t\le 0$ (i.e., its
  initial state is given by the density matrix $\rho \propto
  \exp[-\beta H(0)]$), we use contour-ordered Green functions with
  time arguments on the contour $\CC$ that runs from $t=0$ to $\tmax$
  (the maximum simulated time) on the real axis, back to $t=0$, and
  finally to $t=-i\beta$ along the imaginary axis~\cite{keldyshintro}.
  The local Green function is then given by
  \begin{equation}
    \label{localg}
    G(t,t')
    =
    -i \frac{1}{Z}
    \text{Tr}\big[ e^{-\beta H(0)} \TC \, \hat c(t) \hat c^\mydagger(t')\big],
  \end{equation}
  where $\TC$ is the contour ordering operator, and $\hat
  c^{(\mydagger)}(t)$ are annihilation (creation) operators of the
  mobile particles in the Heisenberg picture with respect to the
  time-dependent Hamiltonian. Up to a factor $i$ the imaginary-time
  Green function of the interacting equilibrium state is recovered
  from Eq.~(\ref{localg}) when both time arguments are on the
  imaginary-time portion of the contour. On the other hand, when both
  time arguments are on the real branch we obtain the
  real-time Green functions $G^\les(t,t')$ $=$ $i \text{Tr}[ e^{-\beta
    H(0)} \hat c^\mydagger(t')\hat c(t)]$ and $G^\lar(t,t')$ $=$ $-i
  \text{Tr}[ e^{-\beta H(0)} \hat c(t) \hat c^\mydagger(t')]$, from
  which various thermodynamic observables can be calculated.  In
  particular, the internal energy per lattice site~(\ref{etot-dmft})
  is given by
  \begin{eqnarray}
    \label{etot-dmft}
    E(t)
    &=
    \partial_t G^\les(t,t')|_{t=t'},
  \end{eqnarray}
  which follows directly from the equations of motion of the lattice
  system, assuming spatial homogeneity.

  The DMFT equations for the Falicov-Kimball model with time-dependent
  interaction were stated in detail in Ref.~\cite{Eckstein2008a} and
  the appendix of Ref.~\cite{Eckstein2008c}, where an analytical
  solution for the case of a sudden switch of the interaction
  parameter is given. The local Green function~(\ref{localg}) is
  determined from the equations of motion~\cite{Eckstein2008a},
  \begin{eqnarray}
    \label{dmft-g}
    G(t,t') = w_0 Q(t,t') + w_1 R(t,t'),
    \\
    \label{dmft-q}
    [i\partial_t + \mu] Q(t,t') - [\Lambda \convz Q](t,t') 
    =\delta_\CC(t,t'),
    \\
    \label{dmft-r}
    [i\partial_t + \mu-U(t)] R(t,t') - [\Lambda \convz R](t,t') 
    =\delta_\CC(t,t'),
  \end{eqnarray}
  where $w_1=1-w_0$ denotes the average number of localized particles
  (which is fixed in the homogeneous phase), and $\Lambda(t,t')$ is
  the coupling to the environment which is obtained by a
  self-consistency condition. The product $[A \convz B](t,t')$
  denotes the convolution of two functions along the contour $\CC$,
  and $\delta_\CC(t,t')$ is the contour delta function.  Throughout
  this paper we use the semielliptic density of
  states~(\ref{rho-semi}), in which case the self-consistency
  condition takes the simple form~\cite{Eckstein2008a}
  \begin{equation}
    \label{selfconsistency-bethe}
    \Lambda(t,t')= V^2 G(t,t').
  \end{equation}

  Equations~(\ref{dmft-g}) to~(\ref{selfconsistency-bethe}) form a
  closed set of integro-differential equation on $\CC$. In this paper
  we consider also the case of a time-dependent hopping amplitude
  in~(\ref{FKM}), i.e., we assume $V_{ij}(t)\equiv V(t) t_{ij}$, where the 
  hopping amplitude $V=V(0)$ sets the energy scale and the 
  $t_{ij}$ are dimensionless. This
  case can be mapped onto the case of an Hamiltonian with
  time-independent hopping and time-dependent interaction in the
  following way: The action $\exp[-i \int dt H(t)]$ of the lattice
  model is invariant under a simultaneous scaling of the Hamiltonian
  $\tilde H(t)$ $=$ $VH(t)/V(t)$ and transformation to new time
  variable $\tilde t(t)$ $=$ $\int_0^t dt' V(t')/V$.  By definition,
  $\tilde H$ has time-dependent interaction $\tilde U(t)= UV/V(t)$ but
  constant hopping $V$, such that Eq.~(\ref{selfconsistency-bethe}) is
  valid.  Under the same transformation of time variables, the
  hybridization function $\Lambda(t,t')$ transforms as $\tilde
  \Lambda(t_1,t_2)$ $=$ $V(t_1)\Lambda(t_1,t_2)V(t_2)/V^2$.  Hence the
  change of the time variable leads to a replacement of the
  self-consistency equation~(\ref{selfconsistency-bethe}) by
  \begin{equation}
    \label{selfconsistency-bethe1}
    \Lambda(t,t')= V(t) G(t,t') V(t').
  \end{equation}

  By discretizing of the contour $\CC$, Eqs.~(\ref{dmft-g}),
  (\ref{dmft-q}),~(\ref{dmft-r}), and~(\ref{selfconsistency-bethe1})
  can in principle be reduced to the inversion of a matrix whose
  dimension is given by the number of mesh points along
  $\CC$~\cite{Freericks2006a,Freericks2008b}. This is approach is not suitable
  here, because it would require a infinite length of the contour in
  case of initial states at zero temperature $(\beta\to\infty)$. A
  different approach first parametrizes contour Green functions in
  terms of various real and imaginary time components, and uses
  Langreth rules to derive separate integral equations for each
  component~\cite{keldyshintro}. One can then remove the imaginary time
  branch by a partial Fourier transform to Matsubara frequencies and
  analytical continuation to real frequencies~\cite{Eckstein2009b}.
  The solution of integro-differential equations such
  as~(\ref{dmft-q}) and~(\ref{dmft-r}) in this way is described in
  detail in Ref.~\cite{Eckstein2009b}. In the following we will
  therefore only briefly restate these equations to mention the
  differences that arise from the fact that we are not solving a
  single equation (such as Eq.~(\ref{dmft-q}) for given $\Lambda$),
  but a nonlinear set of equations.

  In the following we adopt the notation of Ref.~\cite{Eckstein2009b}.
  When both time arguments of a contour Green function $A(t,t')$ are
  on the imaginary time portion of the contour, we obtain the
  Matsubara component which can be represented in the form
  \begin{equation}
    A(-i\tau,-i\tau') 
    =
    \frac{i}{\beta}
    \sum_{n}
    e^{i\omega_n(\tau'-\tau)}
    a(i\omega_n).
  \end{equation}
  The function $a(\omega)$ can be continued to real frequencies.
  Using this parametrization to rewrite Eqs.~(\ref{dmft-q})
  and~(\ref{dmft-r})~\cite{Eckstein2009b}, we obtain the well known
  cubic equations for the local Green function of the homogeneous
  phase in the Falicov-Kimball model~\cite{vanDongen1992a}
  \begin{equation}
    g(\omega+i0) = \frac{w_0}{\omega+\mu-V(0)^2g(\omega+i0)} + \frac{w_1}{\omega+\mu-U-V(0)^2g(\omega+i0)}.
  \end{equation}
  The solution of this equation (with negative imaginary part) describes
  the initial state.

  Next we consider the retarded component
  $A^\ret(t,t')=\Theta(t-t')[A^>(t,t')-A^<(t,t')]$ of the contour
  Green function, for which Eqs.~(\ref{dmft-g}),~(\ref{dmft-q}), (\ref{dmft-r}), and~(\ref{selfconsistency-bethe1})
  read~\cite{Eckstein2009a}
  \begin{eqnarray}
    \label{g-ret}
    G^\ret(t,t') = w_0 Q^\ret(t,t') + w_1 R^\ret(t,t'),
    \\
    \label{q-ret}
    [i\partial_t + \mu] Q^\ret(t,t') - 
    \int_{t'}^t ds V(t)G^\ret(t,s)V(s)Q^\ret(s,t')
    =0,
    \\
    \label{r-ret}
    [i\partial_t + \mu-U] R^\ret(t,t') - 
    \int_{t'}^t ds V(t)G^\ret(t,s)V(s)R^\ret(s,t')
    =0.
  \end{eqnarray}
  These equations must be solved for $t>t'$ with the initial condition
  $G^\ret(t,t)=R^\ret(t,t)=Q^\ret(t,t)=-i$. In contrast to the case
  discussed in Ref.~\cite{Eckstein2009b}, this is a set of {\em
   nonlinear} integro-differential equations. However, the equations
  are causal in the sense that the differential is always determined
  by an integral over the function at {\em earlier} times. Hence the
  solution of Eqs.~(\ref{g-ret}) to~(\ref{q-ret}) is very similar to
  the solution of an ordinary differential equation, and the presence
  of a nonlinearity does not lead to additional difficulties.

  In addition to equations for the retarded and Matsubara components
  one has to consider equations for the mixed components
  $A^{\tv}(t,\tau) \equiv A(t,-i\tau)$.  In terms of the partial 
  Fourier transform
  \begin{equation}
    A^\tv(t,i\omega_n)
    =
    \int_0^\beta d\tau\,
    A^\tv(t,\tau)
    e^{-i\omega_n\tau},
  \end{equation}
  we obtain, after analytically continuing $i\omega_n\to\omega_\pm\equiv\omega\pm i0$,
  \begin{eqnarray}
    \label{g-tv}
    G^\tv(t,\omega_\pm) = w_0 Q^\tv(t,\omega_\pm) + w_1 R^\tv(t,\omega_\pm),
    \\
    \label{q-tv}
    [i\partial_t + \mu] Q^\tv(t,\omega_\pm) - 
    \int_0^t ds V(t)G^\ret(t,s)V(s)Q^\tv(s,\omega_\pm)
    \nonumber\\
    = 
    V(t)G^\tv(t,\omega_\pm)V(0)q^\mats(\omega_\pm),
    \\
    \label{r-tv}
    [i\partial_t + \mu-U] R^\tv(t,\omega_\pm) -
    \int_0^t ds V(t)G^\ret(t,s)V(s)R^\tv(s,\omega_\pm)
    \nonumber\\
    = 
    V(t)G^\tv(t,\omega_\pm)V(0)r^\mats(\omega_\pm),
  \end{eqnarray}
  to be solved with the initial condition
  $A^\tv(0,\omega_\pm)=ia^\mats(\omega_\pm)$.
  Finally, the lesser component satisfies
  \begin{eqnarray}
    \label{g-les}
    G^\les(t,t')
    =
    w_0 Q^\les(t,t') + w_1 R^\les(t,t'),
    \\
    \label{q-les}
    [i\partial_t + \mu] Q^\les(t,t')
    -
    \int_0^t ds V(t)G^\ret(t,s)V(s)Q^\les(s,t')
    \nonumber\\
    =
    -V(t)V(0)
    \int \frac{d\omega}{2\pi} f(\omega) 
    \big[
    G^\tv(t,\omega^+)Q^\vt(\omega^+,t')
    -G^\tv(t,\omega^-)Q^\vt(\omega^-,t')
    \big]
  \nonumber\\
    +
   \int_0^{t'} ds V(t)G^\les(t,s)V(s)Q^\adv(s,t'),
    \\
    \label{r-les}
    [i\partial_t + \mu-U] R^\les(t,t') -
    \int_0^{t} ds V(t)G^\ret(t,s)V(s)R^\les(s,t')
    \nonumber\\
    =
    -V(t)V(0)
    \int \frac{d\omega}{2\pi} f(\omega) 
    \big[
    G^\tv(t,\omega^+)R^\vt(\omega^+,t')
    -G^\tv(t,\omega^-)R^\vt(\omega^-,t')
    \big]
    \nonumber\\
    +\int_0^{t'} ds V(t)G^\les(t,s)V(s)R^\adv(s,t'),
  \end{eqnarray}
  to be solved with the initial condition
  $A^\les(0,t)=A^\vt(0,t)$ for $A=Q,R$.
  Together with the hermitian symmetry, $A^\adv(t,t')=A^\ret(t',t)^*$
  and $A^\tv(t,\omega)=A^\vt(\omega^*,t)^*$ for $Q$, $R$, and $G$ the
  set of equations is closed.

  Note that the expansion of Eqs.~(\ref{dmft-g}),
  (\ref{dmft-q}),~(\ref{dmft-r}), and~(\ref{selfconsistency-bethe1})
  into Eqs.~(\ref{g-ret})-(\ref{r-les}) has the following technical
  advantage.  The numerical solution of Volterra integro-differential
  equations can be easily implemented such that the time
  discretization error scales as $\epsilon$ $\sim$ $\Delta t ^p =
  (\rtime/N)^p$ for $\Delta t\to0$, with $p>1$ \cite{Brunner1986a}.
  To study the excitation energy~(\ref{deltae-tau}) for large $\rtime$
  from the difference~(\ref{exe}), the absolute
  energy~(\ref{etot-dmft}), must be determined with an relative
  accuracy of the order of $\rtime^{-\eta}$. Assuming the above error
  scaling, sufficient accuracy of the energy with respect to $\Delta
  E(\rtime)$ thus requires $N$ $\sim$ $\rtime^{1+\eta/p}$ timesteps
  for $\rtime\to\infty$. On the other hand the Green function must be
  stored at $\mathcal{O}(N^2)$ time points and hence $N$ is the
  limiting numerical factor (we go up to $N$ $\approx$ $10000$). It is
  thus crucial to use an algorithm which is correct up to high-order
  in $\Delta t$ when the exponent $\eta$ is large. For ramps in the
  insulating phase, e.g., $\eta$ $\ge$ $2$ is found
  [cf.~Eq.~(\ref{exe-fkm})], and these results could not be obtained
  using the lowest-order trapezoid approximation ($p$ $=$ $1$), but we
  used higher-order schemes instead ($p$ $=$ $5$).

  \begin{figure}
    \centerline{\includegraphics[width=0.6\columnwidth]{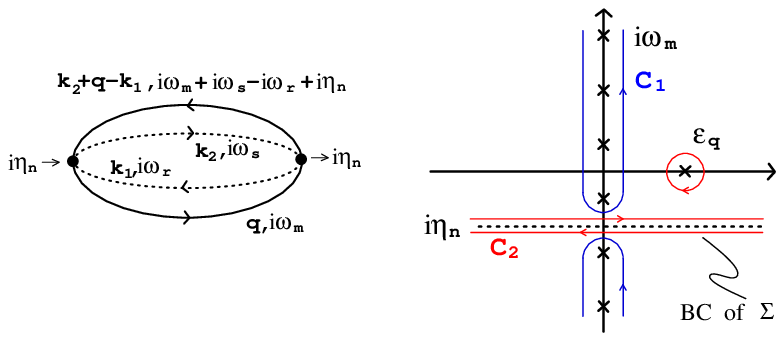}}
    \caption{
      Left panel: Diagrammatic representation of Eq.~(\ref{r2diagram}). 
      Lines represent the noninteracting momentum-resolved Green function
      $g^0_{\Vq\sigma}(i\omega_m)$ $=$ $1/(i\omega_m-\epsilon_{\Vq\sigma})$
      for $\sigma=\uparrow$ (solid lines) and $\sigma=\downarrow$ (dashed lines).
      Momentum is conserved at the vertices, frequency $i\eta_n$ enters at 
      the left vertex. Right panel: Transformation of the Matsubara 
      sum~(\ref{rbosonicmatsubara}) to the real frequency 
      interval~(\ref{rrealfrequency}), using the usual expression
      $\sum_{i\omega_m} w(i\omega_m)$ $=$ $(i\beta/2\pi)\oint_{C_1} 
      dz f(z) w(z)$, where $f(z)$ is the Fermi function and $w(z)$
      is some analytic integrand. The integrand in 
      Eq.~(\ref{rbosonicmatsubara}) has a branch cut at 
      $z= -i\eta_n$ due to the branch cut of
      $\Sigma(z)$ along the real axis, and a pole at 
      $\epsilon_{\Vq\uparrow}$. Then the contour $C_1$ is
      transformed into $C_2$, which yields~(\ref{rbosonicmatsubara}), using 
      that the Fermi function is periodic under shift with 
      bosonic  Matsubara frequencies, $f(\omega-i\eta_n)$ $=$ $f(\omega)$.
    }
    \label{fig-exe-diagram}
  \end{figure}

  \section{Excitation density in the noninteracting limit of the
    Hubbard and Falicov-Kimball model}
  \label{app2}

  For a ramp of the interaction strength in the Hubbard model and the 
  Falicov-Kimball model, the operator in Eq.~(\ref{adi-h}) is given by the 
  double occupation~(\ref{wint}), and the excitation density $R(\omega)$ can 
  be evaluated at $U=0$. For this purpose we introduce the imaginary-time 
  ordered correlation function
  \begin{eqnarray}
    \label{adiR}
    \tilde R(\tau-\tau') 
    = 
    \expval{T_\tau D(\tau) D(\tau')]}_0
  \end{eqnarray}%
  where the expectation value $\expval{\cdot}$ $=$ 
  $\TR[ e^{-\beta H}\cdot]/\TR[ e^{-\beta H_0}]$ is taken in the 
  noninteracting state at temperature $T$ $=$ $1/\beta$ (the limit 
  $T$ $\to$ $0$ is taken at the end), and $D$ is given  by~(\ref{wint}).
  ($T_\tau$ is the imaginary time ordering operator, and $D(\tau)$ are 
  Heisenberg operators with respect to $H_0$). Because $D$ contains an 
  even number of Fermi operators, the function $\tilde R(\tau)$ satisfies 
  periodic boundary conditions on the imaginary time contour $\tau \in [0,\beta]$ 
  and can be expanded in bosonic Matsubara frequencies  $\eta_n$ $=$ $2\pi n/T$,
  $\tilde R(i\eta_n)$ $=$ $\int_0^\beta d\tau \tilde R(\tau) 
  e^{-i\eta_n\tau}$. Using the Lehmann representation one can 
  show that the excitation density~(\ref{fgr-r}) may be obtained from
  the unique analytical continuation of $\tilde R(i\eta_n)$ from $\eta_n>0$ 
  to the upper half of the complex frequency plane,
  \begin{equation}
    \label{r-imag}
    R(\omega) = -\frac{1}{\pi}\text{Im}\;\tilde R(\omega + i0).
  \end{equation}

  To calculate $\tilde R$, the expectation value~(\ref{adiR}) is factorized 
  using Wick's theorem and transformed to bosonic Matsubara frequencies. It 
  turns out that the only nonvanishing contractions for $\eta_n \neq 0$ is 
  given by
  \begin{eqnarray}
    \nonumber
    \tilde R(i\eta_n) 
    &\stackrel{n\neq0}{=}&
    \sum_{\Vk_1,\Vk_2,\Vq} \sum_{r,s,m}
    g^0_{\Vq\downarrow}(i\omega_m)\,\, \times
    \\
    \label{r2diagram}
    && \,\,\times\,\,
    g^0_{\Vk_2\uparrow}(i\omega_s)
    g^0_{\Vk_1\uparrow}(i\omega_r)
    g^0_{\Vk_2-\Vq-\Vk_1,\downarrow}(i\omega_m+i\omega_s-i\omega_r+i\eta_n),
  \end{eqnarray}
  where $g^0_{\Vq\sigma}(i\omega_m)$ $=$ $1/(i\omega_m-\epsilon_{\Vq\sigma})$
  is the noninteracting Green function at momentum $\Vq$, and
  $i\omega_m$ are fermionic Matsubara frequencies.
  The expression has a simple diagrammatic representation (Fig.~\ref{fig-exe-diagram}a). 
  The diagram is split into one Green function line $g^0_{\Vq\downarrow}(i\omega_m)$ 
  and the remainder, which we identify as the second-order contribution 
  $\Sigma^{(2)}_{\Vq\downarrow}$ to the selfenergy,
  \begin{equation}
    \tilde R(i\eta_n) =
    \sum_\Vq \sum_{i\omega_m}
    g^0_{\Vq\downarrow}(i\omega_m) \Sigma^{(2)}_{\Vq\downarrow}(i\eta_n+i\omega_m).
  \end{equation}
  One can now transform the Matsubara summation into a frequency integral,
  where it must be taken into account that the self-energy 
  $\Sigma(z)$ has a branch cut along the real axis with 
  $\Sigma(\omega \pm i0)$ $\equiv$ $\mp \text{Im}\; \Sigma(\omega)$ (Fig.~\ref{fig-exe-diagram}b).
  The result is 
  \begin{equation}
    \label{rbosonicmatsubara}
    \tilde R(i\eta_n) =
    \sum_\Vq \Bigg[
    f(\epsilon_{\Vq\uparrow})
    \Sigma^{(2)}_{\Vq\uparrow}(i\eta_n+\epsilon_{\Vq\uparrow})
    -\frac{1}{\pi}
    \int d\omega\, f(\omega)
    \frac{\text{Im}\;\Sigma^{(2)}_{\Vq\uparrow}(\omega + i0)}{\omega -i\eta_n-\epsilon_{\Vq\uparrow}}
    \Bigg],
  \end{equation}
  where $f(\epsilon)$ is the Fermi function. 
  Eq.~(\ref{rbosonicmatsubara}) is finally continued to the real frequencies by 
  replacing $i\eta_n$ $\to$ $\omega+i0$, and the spectrum~(\ref{r-imag}) is obtained as
  \begin{equation}
    \label{rrealfrequency}
    R(\omega) =
    \frac{1}{\pi}
    \sum_\Vq 
    [f(\omega+\epsilon_{\Vq\downarrow})-f(\epsilon_{\Vq\downarrow})]
    \text{Im}\;\Sigma^{(2)}_{\Vq\downarrow}(\omega+\epsilon_{\Vq\downarrow}).
  \end{equation}
  Taking the limit of zero initial temperature then yields
  Eq.~(\ref{rrealfrequencymain}).

\end{document}